\documentclass[aps,prl]{revtex4}
\def\ee{\end{equation}}
\def\bea{\begin{eqnarray}}
\def\eea{\end{eqnarray}}

\def\ket#1{| #1\rangle}
\def\braket#1#2{\langle \, #1 \, | \, #2 \, \rangle}

\begin{document}

\title{Beable-Guided Quantum Theories: Generalising Quantum Probability Laws}

\author{Adrian \surname{Kent}}
\affiliation{Centre for Quantum Information and Foundations, DAMTP, Centre for
  Mathematical Sciences, University of Cambridge, Wilberforce Road,
  Cambridge, CB3 0WA, U.K.}
\affiliation{Perimeter Institute for Theoretical Physics, 31 Caroline Street North, Waterloo, ON N2L 2Y5, Canada.}
\email{A.P.A.Kent@damtp.cam.ac.uk} 

\date{April 2012} 

\begin{abstract}
We introduce the idea of a {\it beable-guided quantum theory}.
Beable-guided quantum theories (BGQT) are generalisations of quantum theory, 
inspired by Bell's concept of beables. 
They modify the quantum probabilities for some specified set
of fundamental events, histories, or other elements of
quasiclassical reality by probability laws that depend on the
realised configuration of beables.   For example, they may
define an additional
probability weight factor for a beable configuration,
independent of the quantum dynamics. 

BGQT can be fitted to observational data to provide foils against
which to compare explanations based on standard quantum theory.  For
example, a BGQT could, in principle, characterise the effects
attributed to dark energy or dark matter, or any other deviation from
the predictions of standard quantum dynamics, without introducing
extra fields or a cosmological constant.  The complexity of the
beable-guided theory would then parametrise how far we are from a
standard quantum explanation.

Less conservatively, we give reasons for taking suitably
simple beable-guided quantum theories as serious phenomenological
theories in their own right.   Among these are that 
cosmological models defined by BGQT might
in fact fit the empirical data better than any standard quantum
explanation, and that BGQT suggest potentially interesting non-standard ways
of coupling quantum matter to gravity.

\end{abstract}
\maketitle
  
\section{Introduction}

\begin{quotation}
``Considering the pervasive importance of quantum mechanics in 
modern physics, it is odd how rarely one hears of efforts to
test quantum mechanics experimentally with high precision.
$\ldots$ The trouble is that it is very difficult to find any
logically consistent generalization of quantum mechanics.''

(Steven Weinberg, in Testing Quantum Mechanics \cite{weinberg1989testing})
\end{quotation}

Weinberg's paper \cite{weinberg1989testing} exploring ideas about
nonlinear generalizations of quantum theory, from which the quotation
above is taken, led to some fundamental 
insights about the relationship between quantum theory and special
relativity \cite{polchinski1991weinberg,gisin1990weinberg}.
In fact, applied to non-relativistic quantum mechanics,
the quotation was already contradicted by dynamical collapse models
\cite{ghirardi1986unified,GPR}. 
Nonetheless, the belief that generalizations of quantum theory must be 
inconsistent or at least 
suffer from some fundamental sickness remains widespread.  
In particular, the fact that 
Weinberg's nonlinear generalizations of quantum theory
allow superluminal signalling
\cite{polchinski1991weinberg,gisin1990weinberg}
(or, it was argued, inter-universe signalling \cite{polchinski1991weinberg}) 
seem to have persuaded many that relativistic quantum theory,
at least, probably is in some strong sense an 
isolated point in theory space.  Recent no-go theorems
\cite{ck,pbr,cr} may have helped reinforce this impression. 

None of these no-go results actually imply this conclusion, however (nor
do their authors argue that they do). 
And in fact, if one is open-minded about the
mathematical constructions 
one can use, the rules one can postulate, and the ways in which 
quantum laws can be generalized, it is not hard to 
define logically consistent and
potentially scientifically interesting generalizations of quantum
mechanics that do not conflict with special relativity.   
For example, there exists an infinite class of consistent 
nonlinear theories that neither
violate Lorentz invariance nor allow superluminal
signalling \cite{kent2005nonlinearity}.   

This paper describes another infinite class of generalizations of quantum
theory.  This class includes arbitrarily baroque theories, but also includes
subclasses of
relatively simple theories that seem both potentially phenomenologically useful
and interesting in their own right.  
We discuss their potential scientific implications, expanding an earlier
discussion \cite{kent1997beyond} and setting it in a more
general and fundamental context.
   
No specific BGQT are advocated or compared to empirical data here: the 
aim here is to make some simple conceptual points in order to 
expand the boundaries of future research in several directions.

\section{Beable models} 

Arguably \footnote{Some claim that quantum decoherence explains, or
  points towards an explanation, of the appearance of a single
  quasiclassical world following quantum probabilistic laws within a 
many-worlds picture defined by the universal wave function.
For arguments and counterarguments on this point see 
e.g. \cite{mwbook,giulini1996decoherence}; some reasons why I disagree 
can be found in \cite{kentoneworld}} the key problem in quantum foundations -- and one of the key
problems in modern physics -- is the apparent
impossibility of deriving, from unitary quantum theory alone,
an explanation of the appearance of a quasiclassical 
world.\footnote{That is, a world with macroscopic variables following mostly deterministic
classical equations, and also influenced by definite but unpredictable
outcomes of quantum events.}

The most interesting attempts to make progress on this
problem confront it head on, by adding extra 
mathematical structure to unitary quantum theory,
to define what Bell called {\it beables} \cite{bell1976theory, 
bell1987beables}.
The beables are mathematical representations of particle trajectories, or 
processes, or events, or
histories, or whatever the right concept is for the elementary quantities from which we
can build a description of possible quasiclassical worlds (if the idea
works).   
Beable models are indeterministic: they assign a probability measure to configurations of 
beables.   This (if it works) implies a probability measure on possible
quasiclassical worlds, and hence implies probabilistic predictions about the 
quasiclassical world we experience.  In particular, if the beable model is intended
to replicate the predictions of quantum theory precisely, it should be
possible in principle to derive the Born rule for
quantum experiments from the beable configuration probability measure.
So, on this view, our quasiclassical world is the one that 
nature randomly chose to be realized from among all the possible worlds,
and is fundamentally defined by some randomly chosen configuration of
beables from among all the possible configurations.

Examples of beables in beable models or proto-models
include the particle trajectories in de Broglie-Bohm
theories, the collapse centres in discrete dynamical collapse models, 
 and the real history defined by a time-dependent density
matrix in a real world interpretation \cite{akrealworld}.
The consistent or decoherent histories approach \cite{gell1990quantum,griffiths2003consistent} 
can and arguably should \cite{dowker1996consistent,dowker1995properties,kent1998quantum}
also be thought of as a so far unsuccessful attempt at an honest beable
model, in which the beables would be elementary 
histories from a preferred consistent set
defined by an (alas as yet
undiscovered) appropriate quasiclassical set selection
rule.   

\subsection{Beables for sceptics}

\begin{quotation}
``For those people who insist that the only thing that is important
is that the theory agrees with experiment, I would like to imagine
a discussion between a Mayan astronomer and his student.  The Mayans
were able to calculate with great precision predictions, for example,
for eclipses and for the position of the moon in the sky, the
position of Venus, etc.   It was all done by arithmetic.  They
counted a certain number, and subtracted some numbers, and so on.
There was no discussion of what the moon was.  There was no
discussion even of the idea that it went around.  They just 
calculated the time when there would be an eclipse, or when
the moon would rise at the full, and so on.   Suppose that a young
man went to the astronomer and said 'I have an idea. Maybe those
things
are going around, and there are balls of something like rocks out
there, and we could calculate how they move in a completely different
way from just calculating what time they appear in the sky', 
'Yes', says the astronomer, 'and how accurately can you predict
eclipses?'
He says, 'I haven't developed the thing very far yet', Then says
the astronomer, 'Well, we can calculate eclipses more 
accurately than you can with your model, so you must
not pay any attention to your idea because obviously the
mathematical scheme is better'.  There is a very strong 
tendency, when someone comes up with an idea and says, 'Let's
suppose that the world is this way', for people to say
to him, 'What would you get for the answer to such and such
a problem?'  And he says 'I haven't developed it far enough'.
And they say, 'Well, we have already developed it much further, and
we can get the answers very accurately'.   So it is a problem
whether or not to worry about philosophies behind ideas.''

(Richard Feynman, {\it Seeking New Laws} \cite{feynman1967character})
\end{quotation}

Are existing beable models plausibly fundamentally correct? 
Almost certainly not.  Existing beable models look ad hoc, more like 
Heath Robinson \footnote{Or Rube Goldberg, for readers more familiar 
with the American canon.} 
constructions than fundamental physical theories.
They also have some well known problems \footnote{The apparent 
difficulty in constructing beable models that respect Lorentz invariance is one. 
Another is the problem of a double ontology.  Do the beables and the quantum state give apparently equally
valid alternative pictures of reality?  If they do, to claim that
the beable picture resolves any problem you have to be willing to
postulate that {\it they} represent reality, or at least the
reality we experience, and the quantum state, well, just doesn't.
One way of saying this is to say the initial quantum state and
Hamiltonian play law-like roles in the theory, while the beables are
the real physical variables.  Whether this resolves or just restates the
problem is debatable.} -- as (let's not forget) does every
other attempt to date at finding a fundamentally satisfactory understanding of quantum theory.
But we don't necessarily have to find existing beable models
theoretically compelling,
or even believe all their problems can be fixed,
to make scientific use of them.   

Let's assume that the final theory we're heading for is 
as compellingly beautiful as most physicists hope.
Still, it isn't in sight yet.  Why should we be so confident that  
{\it every} step on the path to it involves expressing physical 
insights in terms of mathematically beautiful ideas? 
Maybe a theory unifying quantum theory and gravity, or some other form
of post-quantum theory, 
will emerge in something like the way quantum theory did, from
incomplete ad hoc ideas (like Planck's energy quantization hypothesis),
toy models (like the Bohr atom) and partial insights (like de
Broglie's matter wave hypothesis).  

Especially at the moment, when progress on fundamental problems in physics
is so stalled, it seems a mistake to dismiss ideas that achieve
{\it something}, however unaesthetically.   And beable models {\it do}, after all, despite their problems, give a
logically straightforward way of resolving the tension between classical and quantum
physics -- a tension which has to be resolved {\it somehow}.    
Also in their favour is that -- unlike, for example, attempts to make
sense of many-worlds quantum theory
\cite{kentoneworld} -- they work 
within the only tried and tested scientific paradigm we have, in which
the aim of a scientific theory is to define a single objective reality and 
make standard probabilistic predictions about our observations of that
reality.  Those seem good enough reasons to explore whether beable models
lead to interesting and testable new scientific ideas.   This paper
adds more reasons for believing that they do, and also for believing
that these ideas
may be valuable even if beable models eventually turn out to be only 
rough approximations to a  
deeper theory framed using different concepts.

In summary,  beable models are potentially useful scientific
tools, even for those who query or reject the value 
of beables as a physical concept.   

\section{Beable probabilities in standard beable models}

\subsection{Nonrelativistic beable models}

We start by characterising abstractly a class of beable models of
non-relativistic quantum mechanics, namely {\it time-local} beable
theories.  
In these, 
we suppose we are given quantum theory with some fixed initial state
$\ket{\psi (0) }$ at some initial time $t=0$, and 
some fixed Hamiltonian $H$, which for simplicity and definiteness
we take here to be time-independent.  
Our aim is to describe physics, at the most fundamental level, for
all times $t>0$.  
The possible beable configurations take the form of 
collections $B$ of pairs of beables and
(corresponding) times: 
$$B = \{ ( B_t , t ) : B_t \in \Lambda (t) , t \geq 0
\} \, ,$$
where each $ \Lambda (t)$ is a set of the beables
(some stipulated mathematical quantities) defined
for each $t \geq 0$.  The set
$\Lambda (t)$ may be empty at some or even generic times $t$, and need
not depend continuously on $t$. 

If the model respects standard quantum dynamics for the state vector, it also 
includes the standard time-evolved quantum state $\ket{\psi (t) } = \exp ( - i H t / \hbar )
\ket{ \psi (0) }$ as a mathematical object.   The
quantum state may itself 
also be defined to be one of the beables, but it need not necessarily be.  

Perhaps the most familiar example of a time-local beable model respecting
standard quantum dynamics is standard
non-relativistic de Broglie-Bohm theory \cite{debroglie, bohm}
applied to a system
of $N$ distinguishable particles.  Here the quantum state at time $t$
is $\ket{\psi(t)}$, where
$$
\braket{x_1 , \ldots , x_N}{\psi (t) } = \psi ( x_1 , \ldots , x_N ; t ) \, .
$$
We will take the beables at time $t$ to be the 
position coordinates of the $N$ particle trajectories.  (We will not
take the quantum state here to be a beable, although many
authors choose to.  Both options are possible, and both are
questionable: see footnote $\left[ 40 \right]$ 
above on double ontologies.)   So we have
$$\Lambda(t) = \{ x_1
(t) , \ldots , x_N (t) \} \, . $$    These follow continuous equations,
so in this case $\Lambda(t)$ depends continuously on $t$.

Standard de Broglie-Bohm theory makes experimental predictions indistinguishable
from those of Copenhagen quantum theory where both apply, in the
following sense: we can
always recover the Bohmian predictions for quasiclassical physics 
in the Copenhagen formalism by finding some suitably macroscopic 
system $A$ that undergoes an effectively irreversible measurement
interaction with the measured quantum system $S$, and treating $A$
as though it follows quasiclassical dynamical laws rather than quantum
laws.  However, de Broglie-Bohm theory gives precise 
dynamical equations from which the quasiclassical behaviour of such
objects can be derived, rather than separately postulated.   
De Broglie-Bohm theory can also be straightforwardly be applied to
closed quantum systems and can describe the emergence of
quasiclassical physics within closed systems.
Copenhagen quantum theory, on the other hand, relies on the imprecise
concept of a separate classical realm through which the quantum realm
is probed and measured, and does not apply to closed quantum systems. 

An example of a time-local beable model that respects standard 
quantum dynamics for the state vector but makes distinct predictions from standard
quantum theory is Valentini's modified de Broglie-Bohm theory
\cite{valentini1991signal,valentini2010inflationary},
which follows the de Broglie-Bohm guidance equation but has
a non-standard initial condition on the particle trajectories
at $t=0$: 
$$P ( x_1 (0) , \ldots , x_N (0) ) \neq | \psi ( x_1 , \ldots , x_N ;
0 ) |^2 \, . $$

An example of a time-local beable model that does {\it not} respect standard 
quantum dynamics, and that also makes distinct predictions from standard
quantum theory, is the original Ghirardi-Rimini-Weber discrete 
dynamical collapse model \cite{ghirardi1986unified}.
We consider it here in the form first suggested by
Bell \cite{bell2001there} (see also
e.g. \cite{kent1989quantum,tumulka2006relativistic}), 
where the beables are the (isolated) space-time points about
which the collapses are centred.   
Here, for a system of $N$ distinguishable particles, at generic times
$t$ there are no collapses and hence no beables: $\Lambda(t) = \{ \, \}$. 
 If a collapse centred about
$x$ occurs for particle $i$ at time $t$ then $\Lambda(t) = \{ (x,i)
\}$. 
In cases where  $M \geq 2$ collapses occur at exactly the same time $t$ 
 collapses at time $t$, we have $\Lambda(t) = \{ (x_1 , i_1 ) , \ldots ,
(x_M , i_M ) \}$.  (We include these cases for completeness, although 
the total probability measure for such multiple collapse events, 
integrated over all time, is zero.)
As the GRW model illustrates, $\Lambda(t)$ need not 
necessarily depend continuously on $t$ in a physically
sensible time-local beable model.  

In summary, a time-local beable model defines the possible sets of
time-labelled 
beables,  
$$B = \{ ( B_t , t ) : B_t \in \Lambda (t) , t \geq 0
\} \, , 
$$ 
takes as
input the initial quantum state $\ket{ \psi (0) }$ and the Hamiltonian
$H$ and from these data computes as output 
a probability measure $\mu ( B )$ on the 
sample space of allowed sets $B$.  The measure depends on
$\ket{\psi(0)}$
and $H$: if we think of these as fixed by some particular theory
$T$, we may write $ \mu \equiv \mu_{T}$ to emphasize that the beable
configuration probabilities depend on the (quantum) theory.   

While these familiar examples of beable models are time-local, we
can also imagine more general types of beable model.   For
example, each beable might be associated with an extended region of
space-time.\footnote{A model involving histories of generalized
  quantum events would be an example.}    We can extend the above
characterisation to more general beable models, since nothing in
our abstract definitions relies on time-locality.
Thus, given a theory $T$ defining 
$\ket{ \psi (0) }$ and $H$, a 
{\it beable model} defines the possible sets of beables $ B $
and a probability measure $\mu_{T} ( B  )$. 

\subsection{Relativistic beable models}

Relativistic beable models have proved harder to construct --
perhaps unsurprisingly, given that we have no mathematically
rigorous construction even of non-trivial relativistic quantum 
field theories in $3+1$-dimensional Minkowski space-time.\footnote{ 
There are various proposals for relativistic beable
models (e.g. \cite{tumulka2006relativistic,akrealworld} ).
Assessing these works in
progress is beyond our scope here: our aim is to describe a general class of theories rather
than focussing on specific examples.}

We can straightforwardly extend our abstract characterization
of beable models to quantum field theory in Minkowski
space and quantum cosmology.  This turns out to be useful, despite the
paucity of familiar
concrete examples: see the discussion of coarse-grainings in
cosmological models below.   
   
A Lorentz invariant beable model defines the possible sets of beables $ 
B $ and computes a probability measure $\mu_{T} ( B )$ from a theory
$T$ defining the Hamiltonian $H$ and some 
asymptotic past boundary condition $ \psi_{- \infty} = \lim_{S \rightarrow - \infty} \ket{ \psi_S }$ on
the quantum state associated with spacelike hypersurfaces $S$ tending
to past infinity, {\it by Lorentz covariant rules}.  

Similarly, a generally covariant beable model in quantum cosmology
uses generally covariant rules to define the possible sets of beables $ 
B $ and compute a probability measure $\mu_T (  B )$, given a
generally
covariant theory $T$ defining the quantum evolution law and some 
initial condition postulate (for example, the no-boundary condition).

\section{Beable-guided quantum theory}
\begin{quotation}
``Inert, uninfluential, a simple passenger in the voyage of life, it
is allowed to remain on board, but not to touch the helm or handle the
rigging.''

(William James, {\it Are We Automata?} \cite{james1879we}).
\end{quotation}

There {\it is} something unsettlingly epiphenomenal about the status of the
beables in standard quantum beable
theories.  The quantum state evolution does all the
mathematical work in defining the beable probability distribution; 
the beables, so to speak, hitch a free ride.   

While some non-standard beable
theories give the beables at least a little more of a role, it is still
a secondary one.   

For example, in Valentini's
modified Bohmian theory
\cite{valentini1991signal,valentini2010inflationary}, 
the evolution of the Bohmian particle trajectories throughout time is
determined by the quantum state, just as in ordinary de Broglie-Bohm
theory: the only difference is that the probability distribution of
the initial Bohmian particle positions is chosen independently of the
initial quantum state.   Granted, this is a significant difference,
and has intriguing consequences, but the dynamics remain determined
by the quantum state throughout.  

In GRWP dynamical collapse models \cite{ghirardi1986unified,GPR}, 
there is a genuine interplay between the quantum state and
collapse events: the quantum state at time $t$ depends on
all previous collapse events as well as on the initial state and 
Hamiltonian.   Again, this is a significant generalization of 
standard quantum theory.   
Still, in at least one sense the quantum state still plays a dominant
role: the probability of a collapse taking place at any given
point in Galilean space-time is entirely determined by the quantum
state at that time.  Indeed, all of physics, including the beable
probability 
distributions after time $t$, are determined by the quantum state at
time $t$. 

To be fair, each of these pioneering examples of generalizations of 
quantum theory has its own internal logic that provides  
motivation for the beables playing precisely the role they
do.   Maybe one of these theories, or a theory in which the beables
play a 
similar and similarly secondary role, will indeed turn out to be
a better description of nature than standard non-relativistic quantum
theory.

However, we see a compelling motivation to explore ways of setting the 
beables on a still more equal footing with the quantum state --   
hence the idea of beable-guided quantum theories, to which we now turn. 

\subsection{Beable configuration probability weights and probabilities}

The theory $T$ that 
defines the initial quantum state and Hamiltonian still defines a
probability measure $\mu_{T} ( B )$, as above. However, $\mu_{T} ( B
)$
no longer defines the probability of the beable configuration $B$. 

Instead, we take the actual probability measure of the beable configuration $B$
in a beable-guided quantum theory to be some
function
$$
\mu'_T (B) = f ( \mu_T (B), B ) \, 
$$
that depends on the quantum probability measure and on the beable
configuration.  

This gives a very large class of possibilities indeed.
To be a little more concrete, while still allowing a 
large class of possibilities that includes many interesting
generalizations of quantum theory, in what follows we will
illustrate the idea by 
considering product functions of the form  
$$
f( \mu_T (B) , B) =  C \mu_T ( B ) w( B ) \, , 
$$
where $w(B)$ is a real non-negative weight function. 
The normalised probability measure is then 
$$
\mu'_T (B) =  \mu_T ( B ) w( B )  ( \int_{B'} 
\mu_T ( B' ) w( B' ) )^{-1} \, .  
$$

What defines the weight function $w( B )$?   
We have defined $w(B)$ to be a function only of the beable
configuration $B$, 
independent of the initial quantum state and Hamiltonian,
and to be non-negative and real.  
Modulo these constraints, in principle, {\it any
  rule at all} is allowed. 
We have 
(even in this restricted class) an uncountably infinite class of theories.  
However, in the most interesting cases, the rules defining $w(B)$ 
should be relatively simple.

It is worth stressing again here that the wave function $\ket{
  \psi(t)}$ may be included in the beable set $B$.   If it is, 
a beable-guided quantum theory may be governed by probability 
rules that explicitly depend on the set $\{ \ket{ \psi (t) }
\}_{-\infty < t < \infty} $ or on selected subsets.  Similarly,
relativistic BGQT that include hypersurface-dependent wave
functions $\ket{ \psi (S) }$ as beables may be governed by
probability rules that explicitly depend on $\{ \ket{ \psi (S) }
\}_S$.   

\subsection{Examples of rules for weight functions}

\subsubsection{Simple Bohmian examples}

To give a simple example, we could define a beable-guided quantum theory from a
non-relativistic Bohmian model of two particles with Bohmian
trajectories
$B= \{ x_1 (t), x_2 (t) \}_t$, taking  
$$
w( B ) = \exp ( - ( {\rm lim~sup}_t ( x_1 (t ) - x_2 (t) )^2 / a^2 )) \, . $$ 
This is fairly easy to understand intuitively, although rather ad hoc: 
compared to the standard
Bohmian model, pairs of trajectories are more or less likely to be
selected depending on the closest separation they ever attain. 

A variation that perhaps might appear a little more natural, 
for a model universe of finite duration, from time $0$ to $T$, is 
$$
w( B ) = T^{-1} 
\int_0^{T} dt \exp ( - ( ( x_1 (t ) - x_2 (t) )^2 / a^2 )) \, , $$ 
which prefers pairs of trajectories that stay close over time, with respect
to a simple measure.   
Note however that the limit 
$$
w( B ) = \lim_{ T \rightarrow \infty} T^{-1}
\int_0^{T} dt \exp ( - ( ( x_1 (t ) - x_2 (t) )^2 / a^2 )) \,  $$ 
may not lead to well-defined beable configuration probabilities in
general, since in many examples almost all configurations (with
respect to the standard measure) have $w(B)=0$.\footnote{This model
could be varied by defining $\mu'(B)$ directly via a limit, rather
than $w(B)$.}

To reiterate, any rule at all is allowed in principle.  A more
baroque, less intuitively
interpretable, and presumably correspondingly less interesting example is
given by 

\begin{eqnarray*}
w( B ) &= \alpha \exp ( - ( \int_{0 \leq t \leq 1 } ( x_1 (t )
- x_2 (t) )^2 / a^2 )) + \beta \exp ( - ( \max_{ 2 \leq t \leq 6  } ( x_1 (t )
- x_2 (t) )^2 / b^2 )) \\ & \qquad + \gamma \theta ( \max_{t} (x_1 (t))^2 - c^2 ) + 
\delta \cos^2 ( \max_{t} ( x_1^2 (t) - x_2^2 (t + T ) ))  \, ,
\end{eqnarray*} 
where $\alpha, \beta, \gamma, \delta, a,b,c,T$ are positive
constants and $\theta$ is the Heaviside step function.   

\subsubsection{A simple collapse model example}

A simple beable-guided version of a non-relativistic Ghirardi-Rimini-Weber model for two
distinguishable particles for times $t \geq 0$, with collapse events taking place at
$\{ x^1_i , t^1_i \}_{i=1}^{\infty}$ and $\{ x^2_i , t^2_i
\}_{i=1}^{\infty}$, is defined by 
$$
w( \{ x^1_i , t^1_i \}, \{ x^2_i , t^2_i \} ) = \inf_{i, i'}
\exp ( - ( t^1_i - t^2_{i'} )^2 / T^2 ) \exp ( - (x^1_i - x^2_{i'} )^2 /
X^2 ) \, , 
$$
for constants $X,T$.  Roughly speaking, this tends to favour collapse
event histories that include a pair of collapse events for the two
particles that are nearby in space and time with respect to the scales
$X,T$.  

\section{Coarse-grainings of beables in cosmological models}\label{coarse}

Simple non-relativistic beable-guided quantum theories based on de
Broglie-Bohm theory, the GRW model, or other familiar beable
theories make testably different predictions from standard
quantum theory.   They suggest new ways of parametrising
experimental and observational tests of quantum mechanics.  

However, the class of theories, and so parametrisations,
is infinite.  No one beable-guided
quantum theory leaps out as a clearly more compelling candidate than the rest. 
Also, while intrinsically non-relativistic theories might possibly
still suggest interesting cosmological tests, they are obviously 
fundamentally flawed, and so at best of limited use,
as cosmological models.   Can we get any further?

\subsection{Beable-guided quantum cosmological theories: problems} 

Cosmology poses the sternest test of quantum theory as a 
universal theory, and so seems the likeliest arena
where observation might help select potentially physically relevant
beable-guided quantum theories.
Among the problems are: we don't have a quantum theory
of gravity; we don't have anything approaching a standard quantum
cosmological model 
that starts from a simple theory of initial conditions and fits
all the data; ideas about  Lorentz covariant beable models
are works in progress; ideas about generally covariant beable models
are less substantial still.  

\subsection{Phenomenological BGQTs}

Building a BGQT cosmological theory from fundamental first principles 
may not necessarily be the most fruitful approach.  
To test a cosmological beable model -- or any quantum cosmological
theory -- we don't necessarily need a fine-grained description
of the beables.   The first key test is whether the model explains
(insofar as a probabilistic theory can) the features of the observed
universe.  For this we need to characterize the
possible (mostly) quasiclassical worlds that might be realized in any
given theory -- which we can describe in terms of higher-level
physical quantities that the elementary beables 
must characterize.  In any successful beable quantum theory,
quasiclassical parameters -- the approximate
density of matter in a small region, the average distance between
galaxies, the size of the universe (if finite) at any given
cosmological time -- must be characterized by the beables, at least to
a very good approximation.  In other words, quasiclassical parameters
must, to good approximation, be functions of the elementary beables,
and hence must effectively be higher-order beables.  {\it We can
  define higher-level 
phenomenological beable-guided quantum theories directly in terms of these
  parameters}.

So, insofar as we can talk about quantum
cosmological models at all (and we do, despite all
the theoretical and conceptual
problems) we can also talk about beable-guided quantum cosmological
theories.   If we have covariantly defined quasiclassical parameters,
we can use them to construct covariantly defined beable-guided quantum
cosmological theories.  Theories of this sort were proposed in
Ref. \cite{kent1997beyond}: the present discussion sets them in a more
general and maybe more fundamentally appealing context.

For example, any of the covariant definitions of quasiclassical event
explored in the consistent histories literature might be used to
define a beable-guided cosmological theory.  In particular, we can use
covariant notions of event defined via path integrals
\cite{hartle1991quantum}.  We could, for instance
\cite{kent1997beyond} define quantum cosmologies for an expanding
universe with a cosmological time coordinate in which we stipulate in
advance that, when the compact 3-metric has volume $V_i$, the matter
inhomogeneities are of scale somewhere in the range $(\delta_i^{\rm
  min}, \delta_i^{\rm max} )$, for some sequence $V_1 < V_2 < \ldots <
V_n < \ldots$ of increasing volumes.  More generally, we could
define a probabilistic theory of this type: the probability
distribution for the sequence $\{ \delta_i \}$ is $p ( \{ \delta_i \}
)$.   We can also consider continuous versions of such theories,
defined by appropriate limits.  

Similarly, we can define models for a finite universe that
deterministically or probabilistically constrain the scale of the universe,
or the average separation between galaxies, or any other
quasiclassical quantity, as a function of cosmological time. 
  
Again, of course, this gives us an infinite class of theories,
including arbitrarily baroque ones as well as some quite simple ones.
Collectively, these theories seem ideally designed as foils against
which to test the postulate that initial causes and standard evolution
laws together explain everything that can be explained in physics
\cite{kent1997beyond}.  They can also be used as foils for earlier
non-standard theories, for example in testing the alternative
postulate explored in the two-time cosmology literature, that initial
and final causes, together with the Hamiltonian, suffice.
They are also potentially interesting non-standard theories in their own right.  

\section{Beable-guided quantum theory and gravity}

The problems in unifying quantum theory and gravity are
notoriously deep.  We do not have a consistent
quantum theory of gravity, nor a clear conceptual understanding
of how a picture of macroscopic events taking place in a 
space-time with an apparently relatively well-defined large-scale structure
could emerge from one if we did.  We don't know that a 
quantum theory of gravity (in any conventional sense) is even
what we should be looking for.  

Beable-guided quantum theories suggest a different way of 
thinking about quantum theory and gravity.  Quantum theory
appears to be a good description of the behaviour of matter,
at least at small scales.  The gravitational field appears to
define the structure of a definite associated space-time, at least at
large scales.   Whether the gravitational field is {\it fundamentally} 
quantum or classical or something else, it seems to behave,
at large scales, like a higher-level beable, giving a unique and
definite picture of reality associated with the quantum evolution
of the universe.   Supposing this is correct -- i.e. that 
large-scale features of space-time are defined as higher-level
functions of fundamental beables -- we can use BGQT as a 
framework for defining consistent theories in which the 
gravitational field satisfies interesting constraints.
These constraints need not necessarily arise from standard
expectations  or intuitions about quantum theory and gravity. 
We can also consider constraints that only make sense if one
is looking for a new physical principle embodied in a new 
type of theory, such as a BGQT.   

One example of such a constraint, which is a useful foil against which
to test standard expectations, is to impose by fiat that the gravitational field
must be {\it locally causal}, in a sense that naturally generalizes
Bell's definition of local causality to metric theories \cite{aklocalcausality}.    

Define a {\it past region} in a metric spacetime to be a region which contains
its own causal past, and the {\it domain of dependence} of
a region $R$ in a spacetime $S$ to be the set of points $p$ such that every
endless past causal curve through $p$ intersects $R$.     

Suppose that we have identified a specified past region of spacetime 
$\Lambda$, with specified metric and matter fields, 
and let $\kappa$ be any fixed region with specified metric and matter fields. 

Let $\Lambda'$ be another past region, again with specified metric and
matter fields.  (In the cases we are most interested in, $\Lambda \cap \Lambda'$ 
will be non-empty, and thus necessarily also a past region.)

Define 
$${\rm Prob}( \kappa | \Lambda \perp \Lambda' )$$ 
to be the probability that the domain of dependence of $\Lambda$ 
will be isometric to $\kappa$, given that $\Lambda \cup \Lambda'$ 
form part of space-time, and given that the domains of dependence
of $\Lambda$ and $\Lambda'$ are space-like separated regions.  

Let $\kappa'$ be another fixed region of spacetime with specified metric
and matter fields.  

Define 
$${\rm Prob}( \kappa | \Lambda \perp \Lambda' ; \kappa' )$$
to be the probability that the domain of dependence of $\Lambda$ 
will be isometric to $\kappa$, given that $\Lambda \cup \Lambda'$ 
form part of space-time, that the domain of dependence of 
$\Lambda'$ is isometric to $\kappa'$, and that the domains of 
dependence of $\Lambda$ and $\Lambda'$ are space-like separated. 

We say a metric theory of space-time is {\it locally causal} if
for all such $\Lambda, \Lambda' , \kappa$ and $\kappa'$ the 
relevant conditional probabilities are
defined by the theory and satisfy
$$
{\rm Prob}( \kappa | \Lambda \perp \Lambda' ) 
=  {\rm Prob}( \kappa | \Lambda \perp \Lambda' ; \kappa' ) \, . 
$$

The standard expectation is that our space-time is not locally
causal. 
A Bell experiment in which the measurement outcomes are 
amplified macroscopically so that the gravitational fields in
space-like separated regions depend on the outcomes ought
to produce non-locally causal correlations in the gravitational
fields as well as the measurement outcomes.  A beautiful experiment by
Salart et al. \cite{salart} addressed the related question
\cite{akcausalqt} of whether we can ensure that 
measurements in the two wings of a Bell experiment 
produce definite outcomes in spacelike separated regions,
if we follow the intuitions proposed by Diosi \cite{diosi}
and Penrose \cite{penrose}
and assume that a definite outcome of a quantum measurement
requires a gravitationally macroscopic superposition to
be created.

Salart et al. show the answer is affirmative, if we also
accept Penrose and Diosi's estimates for what constitutes
a gravitationally macroscopic superposition. 
However, the Bell nonlocality of the gravitational field
has not yet been directly tested.  It would be good to 
do so, since there is {\it some} motivation \cite{aklocalcausality} 
for exploring the idea
that the gravitational field might be locally causal, strange 
though such a theory would be, and small though our sliver of doubt on
the point may be.  
It would be easier to see how to write dynamical equations  
for a quasiclassical metric theory -- easier  to see  how
space-time puts itself together from  locally determined
pieces -- if it were locally causal.   
BGQT models incorporating gravity gives a useful way of 
defining models that serve as the requisite foils.   

\section{Discussion}

Any generalization of quantum theory can be seen as a foil for testing
standard theories, a way of parametrizing how well any given
experiment tests the theory, or how well standard quantum 
explanations currently fit observational cosmological data.  
This is certainly sufficient 
motivation for thinking about beable-guided quantum theories.  
Perhaps these and other generalizations of quantum theory
will indeed turn out to be mere foils. 
Perhaps quantum dynamics will indeed survive all experimental tests.
Perhaps some elegant Lorentz and generally covariant  beable extension
of quantum theory and quantum gravity will also explain the appearance
of quasiclassicality and the probabilistic and deterministic laws
governing our quasiclassical  world.  

There is, though, another less conservative motivation for considering
beable-guided quantum theories.  Beables give the best way we have of 
explaining the appearance of quasiclassical physics within a 
quantum world.   But in existing beable models -- even non-standard
models such as those of Valentini and Ghirardi-Rimini-Weber-Pearle -- 
the beables
seem unsettlingly epiphenomenal.  The quantum dynamics do most
or all of the work in defining the beable probability distribution,
yet it is the beables that are 
supposed to represent physical reality, not the quantum state.  
While this is not logically inconsistent, it seems odd that the beables should be 
simultaneously so physically important and so passive.  

Beable-guided quantum theories may not completely eliminate this sense
of unease.  The quantum state still plays the major role, at
least in simple beable-guided theories.   And since quantum theory works
so well, this is a fairly
inescapable feature, at least in theories describing laboratory
experiments.  But they do at least reduce the imbalance:
the beables behave more like self-respecting physical
quantities.   

In short, then, the case for taking beable-guided quantum theories
seriously as fundamental theories in their own right is
that we need beables, and then, once we
have beables, they should play an active role in physics.  

One obvious reason to be sceptical is
that -- with the crucial exception of explaining the appearance
of quasiclassical physics -- standard quantum theory appears to 
work very well.  It seems that any corrections due to 
a beable-guided quantum theory must be very small, and yet
nothing in known physics suggests any obvious reason to expect a small
correction parameter.    But then, similar arguments apply to, for
example, the
cosmological constant, the ratio between gravitational and
electromagnetic force strengths, the degree of parity violation --
and yet the parameters are small in each case.   

There is also a danger of overstating the successes of quantum
theory.  Mainstream cosmological theories tend to assume
quantum theory applies to the universe, and for understandable
reasons.   But we don't actually have a good quantum theory of
gravity, let alone a tested quantum theory of cosmology.  
On cosmological scales, it isn't so clear that
quantum theory {\it does} actually explain all the data well. 

Another obvious criticism is that we have infinitely many
beable-guided quantum theories and no compelling principle for
picking out a small number of them as contenders for
fundamental theories.  If we think of beable-guided quantum theories as 
only stepping stones towards a more compelling post-quantum theory, 
though, this might not be so much of a concern either.  On this
view, perhaps some reasonably simple beable-guided quantum theory will turn
out to be a better theory of nature than quantum theory, but if so, we can only
find out which one empirically, and we will only understand why that
particular beable-guided quantum theory is a good approximation once we have
the deeper post-quantum theory.  

Another interesting speculative possibility is that a guiding condition
of the sort we've explored might be necessary to {\it define} a 
quantum theory of gravity, or to rigorously define physically
relevant relativistic quantum field theories, in the first place.
The idea here is that, by rescaling the probabilities of quantum
events (expressed in terms of beables), and perhaps excluding some
classes of events altogether, guiding conditions could allow
rigorously defined theories to be constructed, although the
underlying unguided quantum theory is not rigorously defined
or even renormalisable.  For example, in principle one could try to 
define a beable-guided quantum field theory that modifies the contributions
to a scattering amplitude so as to remove divergences. 

Note that, even if a BGQT is constructed in a
Lorentz invariant (or generally covariant) way from a similarly
invariant beable theory, it might allow superluminal signalling.
Indeed, we already know that cosmological theories with independent
initial and final boundary conditions can (not surprisingly, given
that they break all standard notions of causality) allow superluminal
signalling \cite{akcausality}.  But as such theories remain
consistent, and evade grandfather paradoxes \cite{akcausality}, it is
not so clear that this should be seen as a disastrous problem.  
It would, in any
case, be very interesting to clarify which types of non-trivial
guidance conditions prohibit superluminal signalling, which allow it
in theory but impose strong practical constraints, and which would
allow it in practice with current technology.

We have presented beable-guided quantum theories in a form that 
perhaps fits most naturally within a block universe picture, in
which reality is defined by one configuration of beables, chosen
randomly from amongst all the possibilities.  To put it picturesquely,
on this view, nature's random choice is made once, before or outside
any physical reality is created, and this choice brings into existence
(in some approximate beable representation) space-time and all events
therein.  

That said, nothing in the definition of beable-guided
quantum theories logically implies any stronger commitment to a block universe
picture than already implied by standard quantum theory.  Although the
appearance of a present time in physics is arguably puzzling in both,
it is not inconsistent with either.  In each case, we can calculate
the probabilities of present or near future events, conditioned on
past events, for successive present times, and so recover a dynamical
picture of beable events randomly happening over time.

\section{Acknowledgements}
I thank Fay Dowker for helpful conversations.
This work was partially supported by a Leverhulme Research Fellowship, a grant
from the John Templeton Foundation, and by Perimeter Institute for Theoretical
Physics. Research at Perimeter Institute is supported by the Government of Canada through Industry Canada and
by the Province of Ontario through the Ministry of Research and Innovation.
\section*{References}

\bibliographystyle{plain}
\bibliography{guided}{}

\begin{thebibliography}{10}

\bibitem{bell1976theory}
J.S. Bell.
\newblock The theory of local beables.
\newblock {\em Epistemological Letters}, 9:11, 1976.

\bibitem{bell1987beables}
J.S. Bell.
\newblock Beables for quantum field theory.
\newblock {\em Quantum implications: Essays in honour of David Bohm}, pages
  227--234, 1987.

\bibitem{bell2001there}
J.S. Bell.
\newblock Are there quantum jumps?
\newblock {\em John S. Bell on the foundations of quantum mechanics}, page 172,
  2001.

\bibitem{bohm}
David Bohm.
\newblock A suggested interpretation of the quantum theory in terms of "hidden"
  variables. i.
\newblock {\em Phys. Rev.}, 85:166--179, Jan 1952.

\bibitem{cr}
R.~Colbeck and R.~Renner.
\newblock Quantum theory cannot be extended.
\newblock {\em Bulletin of the American Physical Society}, 56, 2011.

\bibitem{ck}
J.~Conway and S.~Kochen.
\newblock The free will theorem.
\newblock {\em Foundations of Physics}, 36(10):1441--1473, 2006.

\bibitem{debroglie}
L.~de~Broglie.
\newblock in {S}olvay {C}ongress (1927).
\newblock {\em Electrons et Photons: Rapports et Discussions du Cinquième
  Conseil de Physique tenu à Bruxelles du 24 au 29 Octobre 1927 sous les
  Auspices de l'Institut International de Physique Solvay}, 1928.

\bibitem{diosi}
L.~Diosi.
\newblock Models for universal reduction of macroscopic quantum fluctuations.
\newblock {\em Phys. Rev. A}, 40:1165, 1989.

\bibitem{dowker1995properties}
F.~Dowker and A.~Kent.
\newblock Properties of consistent histories.
\newblock {\em Physical Review Letters}, 75(17):3038--3041, 1995.

\bibitem{dowker1996consistent}
F.~Dowker and A.~Kent.
\newblock On the consistent histories approach to quantum mechanics.
\newblock {\em Journal of Statistical Physics}, 82(5):1575--1646, 1996.

\bibitem{feynman1967character}
R.P. Feynman.
\newblock {\em The character of physical law}, volume~66.
\newblock The MIT press, 1967.

\bibitem{gell1990quantum}
M.~Gell-Mann and J.B. Hartle.
\newblock Quantum mechanics in the light of quantum cosmology.
\newblock {\em Complexity, entropy and the physics of information}, 8, 1990.

\bibitem{ghirardi1986unified}
G.C. Ghirardi, A.~Rimini, and T.~Weber.
\newblock Unified dynamics for microscopic and macroscopic systems.
\newblock {\em Physical Review D}, 34(2):470, 1986.

\bibitem{GPR}
Gian~Carlo Ghirardi, Philip Pearle, and Alberto Rimini.
\newblock Markov processes in {H}ilbert space and continuous spontaneous
  localization of systems of identical particles.
\newblock {\em Phys. Rev. A}, 42:78--89, Jul 1990.

\bibitem{gisin1990weinberg}
N.~Gisin.
\newblock Weinberg's non-linear quantum mechanics and supraluminal
  communications.
\newblock {\em Physics Letters A}, 143(1-2):1--2, 1990.

\bibitem{giulini1996decoherence}
D.~Giulini, E.~Joos, C.~Kiefer, J.~Kupsch, I.O. Stamatescu, and H.D. Zeh.
\newblock Decoherence and the appearance of a classical world in quantum
  theory.
\newblock {\em Decoherence and the appearance of a classical world in quantum
  theory., by Giulini, D.; Joos, E.; Kiefer, C.; Kupsch, J.; Stamatescu, I.-O.;
  Zeh, HD. Springer, Berlin (Germany), 1996, IX+ 366 p., ISBN 3-540-61394-3,
  Price DM 98.00.}, 1, 1996.

\bibitem{griffiths2003consistent}
R.B. Griffiths.
\newblock {\em Consistent quantum theory}.
\newblock Cambridge Univ Pr, 2003.

\bibitem{hartle1991quantum}
J.B. Hartle.
\newblock The quantum mechanics of cosmology.
\newblock In {\em Quantum cosmology and baby universes}, volume~1, pages
  65--157, 1991.

\bibitem{james1879we}
W.~James.
\newblock Are we automata?
\newblock {\em Mind}, pages 1--22, 1879.

\bibitem{kent1989quantum}
A.~Kent.
\newblock ``{Q}uantum jumps''and indistinguishability.
\newblock {\em Modern Physics Letters A}, 4:1839--1845, 1989.

\bibitem{kent1998quantum}
A.~Kent.
\newblock Quantum histories.
\newblock {\em Physica Scripta}, 1998:78, 1998.

\bibitem{kent1997beyond}
A.~Kent.
\newblock Beyond boundary conditions: General cosmological theories.
\newblock In {\em Particle Physics and the Early Universe, Proceedings of
  COSMO-97, L. Roszkowski (ed.)}, pages 562--564. World Scientific, 1998;
  arXiv:0905.0632.

\bibitem{akcausality}
A.~Kent.
\newblock Causality in time-neutral cosmologies.
\newblock {\em Phys.Rev. D}, 59:043505, 1999.

\bibitem{akcausalqt}
A.~Kent.
\newblock Causal quantum theory and the collapse locality loophole.
\newblock {\em Physical Review A}, 72(1):012107, 2005.

\bibitem{kent2005nonlinearity}
A.~Kent.
\newblock Nonlinearity without superluminality.
\newblock {\em Physical Review A}, 72(1):012108, 2005.

\bibitem{aklocalcausality}
A.~Kent.
\newblock A proposed test of the local causality of spacetime.
\newblock {\em Quantum Reality, Relativistic Causality, and Closing the
  Epistemic Circle}, pages 369--378, 2009.

\bibitem{kentoneworld}
A.~Kent.
\newblock One world versus many: the inadequacy of {E}verettian accounts of
  evolution, probability, and scientific confirmation.
\newblock {\em Many worlds?: {E}verett, quantum theory, and reality, Saunders,
  S., Barrett, J., Kent, A. and Wallace, D. (eds.), Oxford Univ. Press}, pages
  307--354; arXiv:0905.0624, 2010.

\bibitem{akrealworld}
A.~Kent.
\newblock Real world interpretations of quantum theory.
\newblock {\em Foundations of Physics}, 42:421--435, (2012).

\bibitem{penrose}
R.~Penrose.
\newblock {\em The Emperor's New Mind}.
\newblock Oxford University Press, 1999.

\bibitem{polchinski1991weinberg}
J.~Polchinski.
\newblock Weinberg's nonlinear quantum mechanics and the
  {E}instein-{P}odolsky-{R}osen paradox.
\newblock {\em Physical Review Letters}, 66(4):397--400, 1991.

\bibitem{pbr}
M.F. Pusey, J.~Barrett, and T.~Rudolph.
\newblock The quantum state cannot be interpreted statistically.
\newblock {\em Arxiv preprint arXiv:1111.3328}, 2011.

\bibitem{salart}
D.~Salart, A.~Baas, JAW Van~Houwelingen, N.~Gisin, and H.~Zbinden.
\newblock Spacelike separation in a {B}ell test assuming gravitationally
  induced collapses.
\newblock {\em Physical review letters}, 100(22):220404, 2008.

\bibitem{mwbook}
S.~Saunders, J.~Barrett, A.~Kent, and D.~Wallace.
\newblock {\em Many worlds?: {E}verett, quantum theory, and reality}.
\newblock Oxford University Press, 2010.

\bibitem{tumulka2006relativistic}
R.~Tumulka.
\newblock A relativistic version of the {G}hirardi--{R}imini--{W}eber model.
\newblock {\em Journal of Statistical Physics}, 125(4):821--840, 2006.

\bibitem{valentini1991signal}
A.~Valentini.
\newblock Signal-locality, uncertainty, and the subquantum {H}-theorem. {I}.
\newblock {\em Physics Letters A}, 156(1-2):5--11, 1991.

\bibitem{valentini2010inflationary}
A.~Valentini.
\newblock Inflationary cosmology as a probe of primordial quantum mechanics.
\newblock {\em Physical Review D}, 82(6):063513, 2010.

\bibitem{weinberg1989testing}
S.~Weinberg.
\newblock Testing quantum mechanics.
\newblock {\em Annals of Physics}, 194(2):336--386, 1989.

\end{thebibliography}

\end{document}